\documentstyle[12pt,twoside,fleqn,espcrc1,psfig]{article}

\newcommand{\AmS}{{\protect\the\textfont2
  A\kern-.1667em\lower.5ex\hbox{M}\kern-.125emS}}
\newcommand{\be}{\begin{equation}}
\newcommand{\ee}{\end{equation}}
\newcommand{\ben}{\begin{eqnarray}}
\newcommand{\een}{\end{eqnarray}}

\newcommand{\msbar}{{\overline {\rm MS}}}

\newcommand{\latt}{{\rm latt}}
\newcommand{\conn}{{\rm conn}}
\newcommand{\disc}{{\rm disc}}
\newcommand{\la}{\langle}
\newcommand{\ra}{\rangle}
\newcommand{\btem}{\bibitem}
\newcommand{\nn}{\nonumber}
\def\simgt{\rlap{\lower 3.5 pt\hbox{$\mathchar \sim$}}\raise 1pt \hbox {$>$}}
\def\simlt{\rlap{\lower 3.5 pt\hbox{$\mathchar \sim$}}\raise 1pt \hbox {$<$}}

\title{Nucleon Spin Structures from Lattice QCD: \\
Flavor Singlet Axial and Tensor Charges}

\author{Yoshinobu Kuramashi \\ \vspace{2mm}
Institute of Particle and Nuclear Studies, \\
         High Energy Accelerator Research Organization(KEK), \\
         Tsukuba, Ibaraki 305, Japan}
       
\begin{document}
\maketitle

\begin{abstract}

The flavor singlet axial and tensor charges of the nucleon
are calculated in lattice QCD. We find 
$\Delta\Sigma=\Delta u+\Delta d+\Delta s=+0.638(54)-0.347(46)-0.109(30)
= +0.18(10)$ for the axial charge and
$\delta\Sigma=\delta u+\delta d+\delta s=+0.839(60)-0.231(55)-0.046(34)
= +0.562(88)$ for the tensor charge.
The result for the axial charge shows reasonable agreement
with the experiment and that for the tensor charge is the first
prediction from lattice QCD before experimental measurements. 

\end{abstract}

\maketitle

\section{Introduction}

The parton spin structure of the nucleon at the twist two
level is characterized by two structure functions
$g_1(x,\mu)$ and $h_1(x,\mu)$ with $x$ being the 
Bjorken variable and $\mu$ being the renormalization
scale (see Ref.~\cite{jj}).  
The functions $g_1$ and $h_1$
represent the quark-helicity distribution and the 
quark-transversity distribution respectively. 
Experimentally the former has been measured by deep
inelastic lepton-hadron scattering(DIS) experiments 
since 1980's. On the other hand, we have no
experimental data for the latter,  which could be measured
in polarized Drell-Yan
processes as well as in inclusive pion or lambda
productions in DIS. The polarized Drell-Yan experiment is 
planned using the relativistic heavy ion collider(RHIC) 
at BNL\cite{imai}.  

The first moment of $g_1$ is called the axial charge
of the nucleon $\Delta q$, which is related to
the nucleon matrix element of the axial vector current
through the operator product expansion.
The flavor singlet component of the axial charge
$\Delta\Sigma=\Delta u+\Delta d+\Delta s$, being
interpreted as the total fraction of proton spin carried 
by quarks in the parton picture, has been a matter of great interest
since results of the European Muon Collaboration experiment indicated 
that $\Delta\Sigma$ is quite small
$\Delta\Sigma=0.12(17)$ and that the
strange quark contribution is unexpectedly 
large $\Delta s=-0.19(6)$\cite{emc}.
Further experiments have since been performed with
proton\cite{smc_1,e143_1,e143_2}, 
deuteron\cite{e143_2,smc_2,e143_3,smc_3} 
and neutron\cite{e142,e154} targets.
Recent global analysis of these experimental data gives
$\Delta \Sigma=0.10^{+0.17}_{-0.11}$ at the renormalization
scale $\mu^2=\infty$\cite{phenom}.

The first moment of $h_1$ is called the tensor charge of the
nucleon $\delta q$, which is written in terms of
the nucleon matrix element of the tensor operator.
It is of great importance to predict
the value of $\delta q$ before experimental measurements.  
We are interested in whether the flavor singlet 
tensor charge $\delta \Sigma=\delta u+\delta d+\delta s$
has also a small value as the flavor singlet axial charge. 

In this report we first give a brief 
description of the calculational method  
for nucleon matrix elements in lattice QCD.
We present our results for  
the axial and tensor charges including both the connected
and the disconnected contributions\cite{axial,tensor}.
For the axial charge our main concern 
is a direct check of the small value of $\Delta\Sigma$
and the large negative contribution of $\Delta s$.  
The result for the tensor charge, which is the first one
from lattice QCD, is compared to that for the axial charge.
The current status of the lattice QCD simulations 
for the flavor singlet nucleon matrix elements is summarized 
in Ref.~\cite{okawa}.  

\section{Formulation of the calculational method 
of nucleon matrix elements}

The axial charge $\Delta q$ and the tensor charge $\delta q$ 
are related to the nucleon matrix elements of the axial
vector current and the tensor operator respectively,
\ben
\la p, s\vert {\bar q}\gamma_\mu\gamma_5 q\vert p, s\ra
&=&2Ms_\mu \Delta q,
\label{eq:def_a} \\
\la p, s \vert \bar{q} i\sigma_{\mu \nu} \gamma_5 q \vert p, s\ra  
&=& 2 (s_{\mu} p_{\nu} - s_{\nu} p_{\mu})\  \delta q , 
\label{eq:def_t}
\een
where $p_\mu$ is the nucleon four momentum, $M$ the
nucleon rest mass and $s_\mu$ the nucleon covariant
spin vector normalized as $s_\mu^2=-1$. 
We will explain how to calculate 
these nucleon matrix elements using lattice QCD. 
Hereafter for the definiteness we consider 
the proton for the nucleon and concentrate on the case of
zero spatial momentum.

We consider a four-dimensional Euclidean space-time lattice
with lattice spacing $a$, where sites are labeled 
by $x\equiv(n_1 a, n_2 a, n_3 a, n_4 a)$ with
$n_1,\cdots,n_4$ integers.
The interpolating field for the proton at site 
$x=(\vec x,t)$ is
constructed to have the correct quantum numbers of the proton,
\be
O_{P_s}(x)\equiv \epsilon^{abc}
\left[\left({}^t u^a(x)C^{-1}\gamma_5 d^b(x)\right)u^c(x)
-\left({}^t d^a(x)C^{-1}\gamma_5
u^b(x)\right)u^c(x)\right], 
\label{eq:O_p}
\ee
where $s$ denotes the proton spinor, $a,b,c$ represent
the color indices and $C=\gamma_4\gamma_2$ is the 
charge conjugation matrix.
The proton mass $m_P$ can be extracted from the exponential
decay of the two-point function projected onto the zero
spatial momentum state, 
\be
C_2(t)=\la 0 \vert  \sum_{\vec x} O_{P_s}({\vec x},t)
\bar O_{P_s}({\vec 0},0)\vert 0 \ra 
\stackrel{{\rm large}\; t}{\longrightarrow}
\frac{\la 0\vert O_{P_s}\vert p,s\ra 
\la p,s\vert {\bar O}_{P_s}\vert 0\ra}{2m_P}{\rm e}^{-m_P t}
\;\;\;{\rm for}\;\;t>0.
\label{eq:2pt} 
\ee
Unwanted contributions of the excited states 
fall off rapidly as $t$ increases,
because their masses are heavier than that of the proton.

The local quark bilinear operator 
on the lattice is defined as
\be
O^\latt_\Gamma(x)\equiv {\bar q}(x)\Gamma q(x)
\;\;\;\;(q=u,d,s)
\ee
where $\Gamma=\gamma_\mu\gamma_5$ $(\mu=1,\cdots,4)$ 
is employed for the axial
charge and $\Gamma=\sigma_{\mu\nu}
=[\gamma_\mu\gamma_\nu-\gamma_\nu\gamma_\mu]/2$ 
$(\mu,\nu=1,\cdots,4)$ for the tensor charge.
The three-point function projected onto the zero spatial
momentum state is used to find the forward proton matrix
element of the operator $O^{\rm latt}_\Gamma$;
\ben
\lefteqn{C_3(t)=\la 0\vert  \sum_{\vec x}
O_{P_s}({\vec x},t)\sum_{{\vec x}^\prime}
O^\latt_\Gamma({{\vec x}^\prime},t^\prime)
\bar O_{P_s}({\vec 0},0)\vert 0 \ra} \label{eq:3pt}  \\
&&\stackrel{{\rm large}\; t-t^\prime,t^\prime}{\longrightarrow}
\frac{\la 0\vert O_{P_s}\vert p,s\ra}
{2m_P}{\rm e}^{-m_P (t-t^\prime)} 
\la p,s\vert O^\latt_\Gamma 
\vert p,s\ra\frac{\la p,s\vert {\bar O}_{P_s}\vert 0\ra}{2m_P}
{\rm e}^{-m_P t^\prime} \;\;\;{\rm for}\;\;t>t^\prime>0.\nn
\een
We define the ratio of the three-point function to the
two-point function, whose time dependence is expected to be
\ben
\lefteqn{R_{O_\Gamma}(t)\equiv
\frac{\la 0\vert \sum_{\vec x}
O_{P_s}({\vec x},t)\sum_{{\vec x}^\prime,t^\prime\neq 0}
O^\latt_\Gamma({{\vec x}^\prime},t^\prime)
\bar O_{P_s}({\vec 0},0)\vert 0 \ra}
{\la 0 \vert \sum_{\vec n}O_{P_s}({\vec n},t)
\bar O_{P_s}({\vec 0},0)\vert 0 \ra}} \label{eq:ratio} \\  
&&\stackrel{{\rm large}\; t-t^\prime,t^\prime}{\longrightarrow}
{\rm const.}
+\frac{\la p,s\vert O^\latt_\Gamma \vert p,s \ra}{2m_P}\,t
\;\;{\rm for}\;\;t>t^\prime>0, \nn 
\een  
where the sum over $t^\prime$ is employed to
increases statistics.
Avoiding the excited state contaminations in small $t$
region
we can extract the matrix element 
$\la p,s\vert O^\latt_\Gamma \vert p,s\ra$ 
by a linear fit of $R_{O_\Gamma}(t)$ in terms of $t$.

We should note that the operator $O^\latt_\Gamma$
defined on the lattice is different from the
operator $O^\conn_\Gamma$ defined in the continuum
regularization scheme, e.g., $\overline{\rm MS}$. 
The operator defined in each regularization scheme can be
related by a renormalization factor $Z_{O_\Gamma}$,
\be
O^\conn_\Gamma(\mu)=Z_{O_\Gamma}(\mu a,\alpha_s) O^\latt_\Gamma(1/a),
\ee
where $\mu$ denotes the renormalization scale and 
$a$ is the lattice spacing.
Using this renormalization factor
we can convert the nucleon matrix element 
on the lattice into that in the continuum
$\overline{\rm MS}$ scheme,
\be
\la p,s\vert O^\conn_\Gamma(\mu) \vert p,s\ra=
Z_{O_\Gamma}(\mu a,\alpha_s)\la p,s\vert 
O^\latt_\Gamma(1/a) \vert p,s\ra.
\ee
$Z_{O_\Gamma}$ can be estimated either perturbatively or
non-perturbatively.
In this report we use the perturbative estimate for 
$Z_{O_\Gamma}$\cite{mz}.

\section{Monte Carlo techniques for calculation 
of two- and three-point functions}

In the path-integral formulation 
the proton two-point function of (\ref{eq:2pt})
is represented by
\ben
C_2(t)=\frac{1}{Z}\int\prod_{y,w,v}\prod_{q} 
dA_\mu(y)d{\bar q}(w)dq(v) 
\sum_{\vec x} O_{P_s}({\vec x},t)\bar O_{P_s}({\vec 0},0) 
{\rm e}^{-S_g\{A_\mu\}-\sum_{q}{\bar q}D_q\{A_\mu\}q},
\label{eq:c2pi}
\een
where $S_g\{A_\mu\}$ is the gauge action on the lattice
and ${\bar q}D_q\{A_\mu\}q$ 
denote the quark action for each flavor $q$. 
Time is rotated into the imaginary axis and $Z$ is the
normalization factor so that $\la0\vert 1\vert 0\ra=1$.
We can evaluate this functional integral
for $A_\mu$, ${\bar q}$ and $q$ using Monte Carlo techniques. 
However, fermions are represented by anti-commuting
Grassmann numbers in the path integral, which we cannot 
manipulate directly with numerical computations.
Therefore we perform the integration for the fermion fields
analytically,
\be
C_2(t)=\frac{1}{Z}\int\prod_{y}dA_\mu(y)
\sum_{\vec x} 
\overline{O_{P_s}({\vec x},t)\bar O_{P_s}({\vec 0},0)}
\prod_{q}{\rm det}D_q\{A_\mu\}
{\rm e}^{-S_g\{A_\mu\}} 
\ee
with the overline meaning the quark field contractions
\be
\sum_{\vec x} 
\overline{O_{P_s}({\vec x},t)\bar O_{P_s}({\vec 0},0)}= 
\epsilon^{abc}\epsilon^{a^\prime b^\prime c^\prime}
G_u^{cc^\prime}(x;0)
{\rm Tr}\{C^{-1}\gamma_5
G_d^{b b^\prime}(x;0)C^{-1}\gamma_5 {}^tG_u^{a a^\prime}(x;0)\} 
+\cdots,
\ee
where $G_q(x;0)$ denotes the quark propagator, which 
is defined as the inverse of the quark matrix 
$D_q\{A_\mu\}$, connecting 
the origin and a site $x=({\vec x},t)$.
Since it takes prohibitively large computer time to evaluate the 
fermion determinant ${\rm det}D_q$ in (\ref{eq:c2pi}) 
we neglect the contribution of
this term putting det$D_q=1$ for all flavor $q$.   
This is the ``quenched approximation'' often employed
in current lattice QCD simulations.  In this approximation
virtual quark loop effects are not taken into account. 

A brief sketch of the calculation of $C_2(t)$ with
the Monte Carlo techniques  
is as follows.

\vspace{2mm}
\noindent
i) an ensemble of
gauge configurations $A_\mu^{(i)}$ $(i=1,\cdots,N_g)$ is generated
with the distribution proportional to 
${\rm exp}\left(-S_g\{A_\mu\}\right)$.

\vspace{2mm}
\noindent
ii) solve the quark matrix inversion on the $i$-th
configuration numerically,
\be
\sum_x D_q\{A_\mu^{(i)}\}(y;x)G_q^{(i)}(x;0)=\delta_{y,0}.
\label{eq:point}
\ee

\noindent
iii) in terms of the quark propagator $G^{(i)}_q(x;0)$ 
the $i$-th proton propagator projected to zero spatial
momentum
$[\sum_{\vec x}\overline{O_{P_s}({\vec x},t)
\bar O_{P_s}({\vec 0},0)}]^{(i)}$ is constructed.

\vspace{2mm}
\noindent
iv) an average over the ensemble gives the vacuum
expectation value,
\be
C_2(t)=\lim_{N_g\rightarrow \infty}\frac{1}{N_g}
\sum_{i=1}^{N_g}\left[\sum_{\vec x} 
\overline{O_{P_s}({\vec x},t)
\bar O_{P_s}({\vec 0},0)}\right]^{(i)}. 
\ee
\vspace{2mm}

\begin{figure}[t]
\centering{
\hskip -0.0cm
\psfig{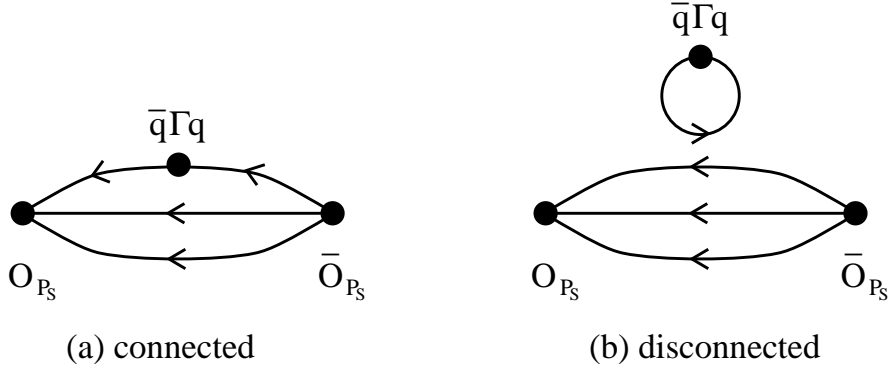}
\vskip -10mm  }
\caption{(a) Connected and (b) disconnected contributions to 
the three-point function of the proton and the operator $O_\Gamma$.} 
\label{fig:sigmadgm}
\vspace{-5mm}
\end{figure}

The three-point function $C_3$ of (\ref{eq:3pt}) has
two types of quark field contractions, represented by a  connected 
[Fig.\ref{fig:sigmadgm}(a)] and a disconnected
[Fig.\ref{fig:sigmadgm}(b)] diagrams.
We should note that in lattice QCD
simulations one can treat both the OZI preserving (connected)
and OZI violating (disconnected) 
contributions even in the quenched approximation.
The calculation of the connected diagram
requires the quark propagator inserted with the
operator $O^\latt_\Gamma$ projected to zero spatial momentum 
at the fixed time slice $t^\prime$;
\be
{\tilde G}_q^{(i)}({\vec x},t;{\vec 0},0)=
\sum_{{\vec x}^\prime}
G^{(i)}_q({\vec x},t;{{\vec x}^\prime},t^\prime)\Gamma
G^{(i)}_q({\vec {x^\prime}},t^\prime;{\vec 0},0).
\ee
We can obtain this quark propagator by
solving\cite{source} 
\be
\sum_x D_q\{A^{(i)}_\mu\}(y;x){\tilde G}^{(i)}_q(x;0)=
\sum_{{\vec x}^\prime}\delta_{y,x^\prime}\Gamma G^{(i)}_q(x^\prime;0).
\ee

There exists a serious difficulty in  the calculation of 
the disconnected piece .
To project the operator with zero spatial momentum
we need the sum of quark loops 
$\sum_{{\vec x}^\prime}
G({{\vec x}^\prime},t^\prime;{{\vec x}^\prime},t^\prime)$
at the time slice $t^\prime$. 
The conventional point source method of (\ref{eq:point})
for each source $({{\vec x}^\prime},t^\prime)$
requires $L^3$ quark matrix inversions, which would take 
about a few months of computer time for each configuration 
if we employ standard parameters used 
in current numerical simulations. 
To overcome this difficulty we use a variant of the method
of wall sources\cite{method}, which makes use of 
local gauge invariance of QCD.
We prepare a quark propagator solved with unit source at
every space site at the time slice $t^\prime$ without gauge fixing:
\be
\sum_x D_q\{A_\mu^{(i)}\}(y;x){\hat G}^{(i)}_q(x)=
\sum_{{\vec x}^\prime}\delta_{y,x^\prime},
\ee
where ${\hat G}^{(i)}_q({\vec x},t)=
\sum_{{\vec x}^\prime}G^{(i)}_q({\vec x},t;
{{\vec x}^{\prime}},t^\prime)$
from the linearity of this equation.
The product of the nucleon propagator and
$\sum_{\vec x}
{\rm Tr}[\Gamma{\hat G}^{(i)}_q({\vec x},{t^\prime})]$
equals the disconnected amplitude up to gauge variant
non-local terms which cancel out in the average over gauge
configurations due to local gauge invariance.
The superior feature of this method is that
it requires only one quark matrix inversion, 
instead of $L^3$ inversions 
for each gauge configuration to calculate 
the spatial sum of quark loops.
For the three-point function in the numerator 
of $R_{O_\Gamma}(t)$ of (\ref{eq:ratio}), 
where we take the sum of the operator over $t^\prime$,
extension of the above calculational 
methods for the connected and
disconnected amplitudes is straightforward. 

\section{Parameters of numerical simulation}

For the lattice action in (\ref{eq:c2pi}) 
we use a single plaquette gauge
action and the Wilson quark action whose parameters are 
$\beta=6/{g^2}$, $g$ being the coupling
constant, and the hopping parameter $K$. The parameter   
$\beta$ controls the lattice spacing $a$ and 
the hopping parameter gives 
the lattice bare quark mass through
$m_q a=(1/K_q-1/K_c)/2$
for each flavor $q$, where 
$K_c$ is the critical hopping parameter at
which the measured pion mass vanishes.

In Table~\ref{tab:param} we summarize our simulation parameters. 
Our calculation is carried out 
in quenched QCD at $\beta=5.7$ on a $16^3\times 20$ lattice.
The up and down quarks are assumed to be
degenerate with $K_u=K_d\equiv K_{ud}$, while the strange quark,
which emerges as the quark loop in the disconnected diagram,
is assigned a different hopping parameter $K_s$. 
We employ three hopping parameters $K_{ud},K_s$=0.160, 0.164
and 0.1665 corresponding to the physical pion mass $m_\pi
\approx 970$, 720 and 520MeV respectively, where the lattice
spacing $a$ is estimated from $m_\rho a$ in the chiral limit
using $m_\rho=770$MeV. 
The physical spatial size $La$ is approximately $2.2$fm.
We analyzed 260 gauge configurations for the
axial charge and 1053 configurations for the tensor charge,
which are generated by the  pseudo heat bath sweeps and are separated by 
1000 sweeps.  In order to avoid
contamination from the negative-parity partner of the proton
propagating backward in time, we employ the Dirichlet
boundary condition in the temporal direction for quark
propagators. We fix gauge configurations on the $t=0$ time
slice, where the proton sources are placed, 
to the Coulomb gauge to enhance proton signals.
Statistical errors for all measured quantities are estimated
by the single elimination jackknife procedure. 

The strong coupling constant in the 
$\msbar$ scheme is evaluated from
the tadpole-improved one-loop relation between the
$\msbar$ coupling and the lattice bare coupling,
$\alpha_\msbar(\pi/a)^{-1}=P_{\rm av}\alpha_\latt^{-1}+0.30928$
with $P_{\rm av}$ the Monte Carlo expectation value 
of the plaquette\cite{alpha}. Reducing the scale from $\pi/a$ to $1/a$ 
by two-loop running we obtain $\alpha_\msbar(1/a)$, 
which is used for the calculation of the
perturbative renormalization factors. 
An estimate of physical nucleon matrix elements
requires the physical value of the degenerate up and down quark mass
$m_u a=m_d a\equiv m_{ud} a$ 
and that of the strange quark mass $m_s a$.
From fits of the measured meson mass spectrum we find
$(m_\pi a)^2=B_{PS}m_{ud} a$ and 
$m_\rho a=A_V+B_V m_{ud} a$.
Using the physical $\pi$ to $\rho$ mass ratio 
$m_\pi/m_\rho=0.18$ we obtain the degenerate up and down quark
mass $m_{ud} a$.
The strange quark mass is estimated by generalizing the 
relation $(m_\pi a)^2=B_{PS}m_{ud} a$ to 
$(m_K a)^2=B_{PS}(m_{ud} a+m_s a)/2$ and using the ratio
$m_K/m_\rho=0.64$. 

\begin{table}[t]
\vspace{-4mm}
\begin{center}
\caption{\label{tab:param}
Simulation parameters for the calculation of
(a) the axial charge and (b) the tensor charge.
$m_N a$ is the nucleon mass in the chiral limit.}
\vspace*{2mm}
\begin{tabular}{llllllll}
\hline
&\multicolumn{7}{l}{$\beta=5.7$, 
$L^3\times T=16^3\times 20$, $K=0.16,0.164,0.1665$} \\
 & \#conf. & $K_c$ & $a^{-1}$(GeV) & $m_N a$ 
& $\alpha_\msbar(1/a)$ & $m_{ud} a$ & $m_s a$ \\
\hline
(a) & 260 & 0.16941(5) & 1.458(18)  
& 0.773(13) & 0.2207 & 0.00340(9) & 0.0826(22) \\
(b) & 1053 & 0.16925(3) & 1.418(9) 
& 0.8024(65) & 0.2158 & 0.00347(5) & 0.0896(12) \\
\hline
\end{tabular} 
\end{center}
\vspace{-7mm}
\end{table}

\section{Results}

\subsection{Axial charges}

We calculate the ratio of (\ref{eq:ratio}) 
employing the axial vector current 
$A_q={\bar q}\gamma_\mu\gamma_5 q$ $(q=u,d,s)$. 
In Fig.~\ref{fig:rl16k164_a}(a) 
we plot the connected contribution of $u$ and $d$ quarks to the 
ratio $R_{A_q}(t)$ for the case of $K_{ud}=0.164$.
A clear linear behavior is observed up to $t\approx 14$ 
for both contributions.
For the disconnected contribution the quality of our data
is not very good, in spite of high statistics of the simulation
(see Fig.~\ref{fig:rl16k164_a}(b)).
The region showing a linear dependence is limited to
$t\approx 5-10$, and errors grow rapidly with increasing $t$;
for $t\ge 12$ the signals are lost in the large noise.
Nevertheless, we can still observe that 
the slope of the ratio for the disconnected contribution 
is smaller than  that for
the connected contribution and it has a negative value. 
To extract the axial charge for the connected and
disconnected contributions 
we fit the data for $R_{A_q}(t)$ to the
linear form (\ref{eq:ratio}) with the fitting range chosen
to be $5\le t\le 10$. 

The axial charges defined on the lattice are 
converted to those in the
continuum $\msbar$ scheme using the tadpole-improved
renormalization factor\cite{mz,lm}
\be
Z_{A_q}(\alpha_\msbar(1/a))=
\left(1-\frac{3K_q}{4K_c}\right)\left[1-0.31 \alpha_\msbar(1/a)\right].
\label{eq:z_a}
\ee
The results for $\Delta u$, $\Delta d$ and $\Delta s$
in the $\msbar$ scheme are summarized in Table~\ref{tab:axial}. 
We should note that the flavor singlet axial vector current
requires an additional lattice-to-continuum divergent
renormalization from diagrams containing the triangle
anomaly diagram. We leave out this factor, since the
explicit form of this contribution which starts at two-loop
order has not been computed yet. 

\begin{figure}[htb]
\begin{minipage}[t]{80mm}
\centering{
\hskip -0.0cm
\psfig{file=rl16k164_a.epsf,width=73mm,angle=0}
\vskip -10mm  }
\caption{Connected and disconnected contributions to $R_{A_q}(t)$
for $u$ and $d$ quarks  at $K_{ud}=0.164$. Solid lines are
linear fits over $5\le t\le 10$.} 
\label{fig:rl16k164_a}
\end{minipage}
\hspace{\fill}
\begin{minipage}[t]{75mm}
\centering{
\hskip -0.0cm
\psfig{file=delud_a.epsf,width=73mm,angle=-90}
\vskip -10mm  }
\caption{Axial charges for $u$ and $d$ quarks as a function
of degenerate $u$ and $d$ quark mass. Open symbols denote
the values extrapolated linearly to the chiral limit.} 
\label{fig:delud_a}
\end{minipage}
\vspace{-5mm}
\end{figure}

\begin{table*}[b]
\vspace{-7mm}
\begin{center}
\caption{\label{tab:axial}
Proton axial charges in the $\overline{\rm MS}$ scheme 
as a function of $K_{ud}$. $\Delta d_\disc$ equals $\Delta u_\disc$.
Values at $K_c$ are obtained by a linear fit in $1/K$.}
\vspace*{2mm}
\begin{tabular*}{\textwidth}{@{}l@{\extracolsep{\fill}}llllll}\hline
$K_{ud}$\hspace{10mm} & $\Delta u_\conn$ & $\Delta d_\conn$ &
$\Delta u_\disc$ & \multicolumn{3}{c}{$\Delta s_\disc$\ } \\
\hline
&&&& $K_s$=0.1665 & 0.1640 & 0.1600 \\
0.1600 & $0.9071(92)$ & $-0.2470(35)$ & $-0.025(10)$ & $-0.025(10)$ & $-0.035(12
)$ & $-0.042(15)$ \\
0.1640 & $0.839(19)$  & $-0.2382(87)$ & $-0.066(23)$ & $-0.049(19)$ & $-0.066(23
)$ & $-0.076(27)$ \\
0.1665 & $0.818(39)$  & $-0.231(23)$  & $-0.093(54)$ & $-0.068(35)$ & $-0.084(46
)$ & $-0.093(54)$ \\
$K_c$  & $0.763(35)$  & $-0.226(17)$  & $-0.119(44)$ &&& \\
\hline
\end{tabular*} 
\end{center}
\vspace{-7mm}
\end{table*}

We present the axial charges for $u$, $d$ and $s$ quarks  
in Fig.~\ref{fig:delud_a} as a function of the 
bare quark mass $m_{ud}$ in physical units. 
As we already remarked the values of the disconnected 
contribution(circles) are small and negative. 
Their magnitude increases slightly toward the chiral limit,
while the connected contributions decrease.

We calculate the physical values of matrix elements
in the following way.  
For $u$ and $d$ quarks, we
estimate the sum of disconnected and connected contributions 
by first combining the two
contributions in the ratio $R_{A_q}(t)$ and then fitting the result 
to the linear form ({\ref{eq:ratio}})
over $5\leq t\leq 10$ for each $K_{ud}$. 
The fitted values are extrapolated linearly to the
chiral limit $m_{ud}=0$, where we neglect the degenerate
$u$ and $d$ quark mass (see Table~\ref{tab:param}).
For the strange quark contribution 
similar extrapolations to $m_{ud}=0$ are made
for each $K_s$, and their results in turn are 
interpolated to the strange quark mass $m_s$. 
This analysis yields for the
quark contribution to proton spin, 
\be
\Delta\Sigma=\Delta u+\Delta d+\Delta s 
=+0.638(54)-0.347(46)-0.109(30) 
= +0.18(10). 
\label{eq:result_a}
\ee
These values, notably the sign and magnitude 
of the strange quark contribution, show a reasonable
agreement with the phenomenological estimate 
$\Delta \Sigma=0.10^{+0.17}_{-0.11}$\cite{phenom}.

Possible sources of systematic errors in our results are
scaling violation effects due to a fairly large lattice
spacing $a\approx 0.14$fm at $\beta=5.7$, 
quenched approximation 
and uncertainties in the perturbative estimate of
the renormalization factor (\ref{eq:z_a}).
For the first two systematic errors we can roughly estimate 
its magnitude from our result of the flavor non-singlet
axial charge  $g_A=\Delta u-\Delta d=0.985(25)$ at $m_{ud}=0$, 
which is about $25\%$ smaller
than the experimental value $g_A=1.2601(25)$\cite{pdg}.
The small value of $g_A$, possibly arising from these
uncertainties, suggest that our result for $\Delta\Sigma$
might be underestimating the continuum value by a similar
magnitude.
For the perturbative renormalization factor
we note that the lack of two-loop calculation for the flavor 
singlet lattice-to-continuum renormalization factor makes it 
difficult to specify the scale at which $\Delta\Sigma$ is
evaluated, although we expect the scale dependence 
to be weak, being a two-loop effect.
These points should be examined in future studies.


\subsection{Tensor charges}

So far, there exist several model calculations of the tensor charge.  
Non-relativistic quark model predicts
$ \delta u  =  \Delta u = 4/3$ and 
$ \delta d  =  \Delta d = -1/3$, while relativistic quark wave functions
with non-vanishing lower components
lead to $\delta u = 1.17$ and $\delta d = -0.29$ together with the
inequality
$\vert \delta q \vert > \vert \Delta q \vert $ $(q=u,d)$\cite{HJ}.
There also exist attempts to calculate $\delta q$ using  QCD sum rules
$(\delta u=1.33(53), \delta d=0.04(2))$\cite{HJ,ioffe}
and a chiral quark model $(\delta u=1.12, \delta d=-0.42, 
\delta s=-0.008)$\cite{goeke}.
The main deficiencies of these model calculations are as follows.

\vspace{2mm}
\noindent
i) the renormalization scale where the matrix
elements are evaluated is not clear.
The tensor current has an anomalous dimension
at the one-loop level, while that of the flavor singlet axial
current starts from the two-loop level.

\vspace{2mm}
\noindent
ii) it is hard to estimate the contribution of OZI violating process,
especially strange quark contributions, 
in a reliable manner. 

\vspace{2mm}
\noindent
Lattice QCD calculation is free from these problems
even in the quenched approximation.

\begin{figure}[htb]
\begin{minipage}[t]{80mm}
\centering{
\hskip -0.0cm
\psfig{file=rl16k164_t.epsf,width=73mm,angle=0}
\vskip -10mm  }
\caption{Same as Fig.~\protect{\ref{fig:rl16k164_a}} 
for $R_{T_q}(t)$. Solid lines are
linear fits over $6\le t\le 11$.} 
\label{fig:rl16k164_t}
\end{minipage}
\hspace{\fill}
\begin{minipage}[t]{75mm}
\centering{
\hskip -0.0cm
\psfig{file=delud_t.epsf,width=73mm,angle=-90}
\vskip -10mm  }
\caption{Same as Fig.~\protect{\ref{fig:delud_a}} for
the tensor charge.} 
\label{fig:delud_t}
\end{minipage}
\vspace{-5mm}
\end{figure}

We calculate the ratio of (\ref{eq:ratio})
for the tensor operator 
$T_q={\bar q}\sigma_{\mu\nu} q$ $(q=u,d,s)$.
The ratio $R_{T_q}(t)$ are plotted 
in Fig.~\ref{fig:rl16k164_t} for the case of $K_{ud}=0.164$.
Fig.~\ref{fig:rl16k164_t}(a)  
shows a clear linear behavior 
for the connected  contributions  
up to $ t \sim 13$, beyond which errors grow rapidly.
For the disconnected contribution in
Fig.~\ref{fig:rl16k164_t}(b)  the data stay   
around zero with $100\%$ errors and
does not show clear signal of a linear dependence on $t$. 
We extract the tensor charge 
both for the connected and disconnected contributions 
fitting the data of $R_{T_q}(t)$
with a linear form (\ref{eq:ratio}) over $6\le t\le 11$.

The results for $\delta q$ are corrected by the 
one-loop renormalization factor\cite{mz,lm}  
\be
Z_{T_q}(\mu a,\alpha_\msbar(1/a))
=\left( 1- {3 K_q \over 4 K_c} \right)
\left[  1 - \left({2 \over 3\pi} \ln (\mu a) + 0.44\right)\  
\alpha_\msbar(1/a) \right].
\label{eq:z_t}
\ee
The tensor charges $\delta u$, $\delta d$ and $\delta s$
in the $\msbar$ scheme are tabulated in Table~\ref{tab:tensor}. 

In Fig.~\ref{fig:delud_t} we plot the tensor charges 
as a function of the lattice bare quark mass $m_{ud}$.
The connected contributions for $u$ and 
$d$ quarks decreases in magnitude as the quark mass decreases.
This trend is the same as for the axial charge.
For the disconnected contribution our result is consistent
with zero, which indicates that the 
OZI violating effects are negligible for the tensor charge. 

\begin{table*}[b]
\vspace{-7mm}
\begin{center}
\caption{\label{tab:tensor}
Proton tensor charges at $\mu = 1/a$ in the
$\overline{\rm MS}$ scheme as a function of $K_{ud}$.
$\delta d_\disc$ equals $\delta u_\disc$.
Values at $K_c$ are obtained by a linear fit in $1/K$.}
\vspace*{2mm}
\begin{tabular*}{\textwidth}{@{}l@{\extracolsep{\fill}}llllll}\hline
$K_{ud}$\hspace{10mm} & $\delta u_\conn$ & $\delta d_\conn$ &
$\delta u_\disc$ & \multicolumn{3}{c}{$\delta s_\disc$\ } \\
\hline
&&&& $K_s$=0.1665 & 0.1640 & 0.1600 \\
0.1600 & 1.072(50) & $-0.251(21)$ & $-0.008(12)$ & $-0.022(16)$ &$-0.013(14)$
& $-0.008(12)$ \\
0.1640 & 0.994(11) & $-0.220(52)$ & $-0.027(28)$ & $-0.041(33)$ &$-0.027(28)$
& $-0.013(25)$ \\
0.1665 & 0.948(31) & $-0.196(15)$ & $-0.044(69)$ & $-0.044(69)$ &$-0.035(58)$
& $-0.009(51)$ \\
$K_c$  & 0.893(22) & $-0.180(10)$ & $-0.054(54)$ &&& \\
\hline
\end{tabular*} 
\end{center}
\vspace{-7mm}
\end{table*}

To extract the physical tensor charge for $u$ and $d$ quarks 
the fitted values are linearly extrapolated to the
chiral limit $m_{ud}=0$.
For the strange quark contribution we first make linear interpolation 
to the physical strange quark mass $m_s a$ at each 
fixed value of $K_{ud}$ for nucleon, 
and then the results are linearly extrapolated to 
the chiral limit.
The final results for the tensor charge is
\be
\delta\Sigma=\delta u+\delta d+\delta s 
=+0.839(60)-0.231(55)-0.046(34) 
= +0.562(88), 
\label{eq:result_t}
\ee
where the error of 
of the sum of $u$, $d$, $s$ contributions are estimated by
quadrature.

Due to the smallness of the disconnected contributions,
the flavor singlet tensor charge $\delta \Sigma$ 
is not much suppressed from its
quark model value. This is in contrast to the
flavor singlet axial charge $\Delta \Sigma$ which
suffers from large suppression.
The ratio $\vert \delta q/\Delta q\vert$ extrapolated to the 
chiral limit gives $\vert \delta u/\Delta u\vert=1.32(15)$
and $\vert \delta d/\Delta d\vert=0.67(18)$, which are 
quantitative different from the prediction of the
nonrelativistic quark model $\vert \delta q/\Delta q\vert=1$ 
$(q=u,d)$ and also have qualitative difference
from the prediction of relativistic quark models
$\vert \delta q/\Delta q\vert> 1$ $(q=u,d)$ mentioned before.
The smallness of $\delta q_\disc$ 
could be related to the C (charge conjugation)-odd and
chiral-odd  nature
of the tensor operator $\bar{q} \sigma_{\mu \nu} \gamma_5 q$ \cite{jt}.

Possible systematic errors originates from
the scaling violation effects
and the quenched approximation.
We may estimate that the magnitude of these systematic
errors is $\sim 20\%$ as for the axial charges.
Toward a definitive determination of the tensor charge
a repetition of the calculation
including dynamical quark effects
with smaller lattice spacing
is required, which we leave for future investigations. 

%

\section*{Acknowledgement}
 
I would like to thank my collaborators S.~Aoki, M.~Doui, 
M.~Fukugita, T.~Hatsuda, M. Okawa and A. Ukawa for a
pleasant joint venture. I am grateful to A.~Ukawa
for valuable comments and careful reading of the manuscript.


\end{document}